# Characterize the non-Gaussian diffusion property of cerebrospinal fluid using Diffusion Kurtosis Imaging and explore its diagnostic efficacy for Alzheimer's disease

Yingnan Xue, MS,[1] Min Wen, MD,[2] Qiong Ye, PhD[3]

[1]Department of Radiology, The First Affiliated Hospital of Wenzhou Medical University, Wenzhou, China
[2]Department of Neurology, Guangzhou First People's Hospital, Guangzhou, China
[3]High Magnetic Field Laboratory, Hefei Institutes of Physical Science, Chinese Academy of Sciences, Hefei, China

**Corresponding author:**

1. Name：Qiong Ye

2. Address: 350 Shushanhu Road, Hefei 230031, Anhui, P. R. China

3. Telephone: +0086 13600676534

4. E-mail address: qiong.ye@hmfl.ac.cn

**Acknowledgments：**

This work was supported by the Collaborative Innovation Program of Hefei Science Center, CAS (2020HSC-CIP010) and Collaborative Innovation Key Foundation of Hefei Science Center, CAS (2022HSC-CIP003).

**Running Title：Non-Gaussian diffusivity of CSF**

# Characterize the non-Gaussian diffusion property of cerebrospinal fluid using Diffusion Kurtosis Imaging and explore its diagnostic efficacy for Alzheimer's disease

## Abstract

Differentiating Alzheimer's disease (AD) patients from healthy controls (HC) remains a challenge. The changes of protein level in cerebrospinal fluid (CSF) of AD patients have been reported in the literature. Macromolecules will hinder the movement of water in CSF and lead to non-Gaussian diffusion. Diffusion kurtosis imaging (DKI) is a commonly used technique for quantifying non-Gaussian diffusivity. In this study, we used DKI to evaluate the non-Gaussian diffusion of CSF in AD patients and HC. Between-group difference was explored. In addition, we have built a prediction model using cross-validation Support Vector Machines (SVM), and achieved excellent performance. The validated area under the receiver operating characteristic curve(AUC) is in the range of 0.96-1.00, and the correct prediction is in the range of 87.1% - 90.0%.

**Introduction**

Alzheimer's disease (AD) is a neurodegenerative disease that usually occurs in the elderly. The main clinical symptoms are progressive cognitive and behavioral impairment. Its etiology is still unclear, mainly related to genetic and environmental factors, and cannot be cured at present. Comprehensive treatment can alleviate the symptoms and delay the progression. The current gold standard for the definitive diagnosis of AD is the identification of disease defined regional patterns of neurofibrillary tau tangles and amyloid plaque malformations in the brain by postmortem examination(Arnold et al., 1991; Braak and Braak, 1991; Brun and Gustafson, 1976; Hardy, 2006; Hyman et al., 2012; Montine et al., 2012; Selkoe and Hardy, 2016). It is still a challenge to distinguish non-invasive or minimally invasive between AD patients and HCs.

Several previous studies have proved the sensitivity of diffusion kurtosis imaging (DKI) index to the microstructure changes of white matter and gray matter, and discussed its diagnostic value for AD(Jensen et al., 2005; Tu et al., 2021; Xue et al., 2019). In biological tissues and organs, the diffusion of water molecules is hindered by membranes and macromolecules. Therefore, the diffusivity of water molecules will deviate from Gaussian diffusion. DKI can provide the non-Gaussian diffusion degree of water molecules in biological tissues and uniform solutions(Jensen et al., 2005; Rosenkrantz et al., 2015). However, the potential of DKI index of CSF in the diagnosis of AD has not been explored previously. As a uniform solution, CSF is more suitable for analysis with DKI than brain parenchyma. Amyloid-β 1-42 ($A\beta_{1-42}$) and total-tau (tau) are well-accepted CSF biomarkers of AD(Tardif et al., 2018). Altered $A\beta$ and tau levels in CSF of AD patients have been reported in several literature(Blennow and Zetterberg, 2018; Falgàs et al., 2020; Shahid et al., 2022). The change of chemical composition in CSF may change its non-Gaussian diffusion characteristics.

In our work, a scalar kurtosis quantity was derived from diffusion-weighted tracer signal averaged over three orthogonal diffusion directions. Then, the non-Gaussian diffusion property in CSF of AD patients and HCs was investigated using DKI. The group comparison was performed, followed by risk modeling to evaluate its diagnostic power. Cross-validation was applied in risk modeling to avoid over-fitting.

**Materials and Methods**

Participants

This study was approved by the institutional review board of the hospital. All participants provided written informed consent. There were fifty-one subjects in this prospective study, with thirty-one patients who were diagnosed with probable AD and twenty HC subjects without cognitive dysfunction. The inclusion criteria for patients with AD were as follows: (1) Age in the range of 50–85 years; (2) Completion of the neurological tests; (3) Fulfillment of the National Institute of Neurological and Communicative Disorders and Stroke/Alzheimer's Disease and Related Disorder Association (NINCDS/ADRDA) criteria (McKhann et al., 1984) for probable AD; (4) $11 \leq$ MMSE score $\leq 22$ (primary school education), $11 \leq$ MMSE score $\leq 26$ (higher than middle school education); (5) A Hachinski Ischemia Score $\leq 4$; (6) A Hamilton Depression Rating Scale (HAMD17) $\leq 10$;

(7) Memory lapses for at least 12 months with a progressively worsening trend; (8) Fazekas scale for WM lesions $\leq 1$ (age $\leq 70$ years), Fazekas scale for WM lesions $\leq 2$ (age $> 70$ years). All subjects were required to have two or fewer lacunar ischemia strokes (of diameter $> 2$ cm) in the brain, as revealed by FLAIR; (9) Neurological examination showed no positive signs; (10) Patients had no non-AD dementias, psychiatric disorders, Parkinson's disease, neuromyelitis optica, seizures, brain trauma, tumors, stroke, etc., that could compromise their cognition; (11) Identification of a responsible and consistent caregiver. The HC group did not have any history of cognitive decline, neurological disorders, or uncontrolled systemic medical disorders.

MRI acquisition

All MRI was examined on a 3.0 T clinical MRI system (Achieva, Philips, Best, Netherlands) with an 8-channel head coil. Multiple b-values DWI images were acquired with 14 b-values (b = 0, 25, 50, 75, 100, 150, 200, 500, 800, 1000, 2000, 3000, 4000, 5000 s/mm$^2$) using a single-shot diffusion-weighted spin-echo EPI sequence. The other parameters of multiple b-values DWI were as follows: repetition time (TR)/ echo time (TE) = 8000/113 ms, flip angle = 90°, slice thickness = 5 mm without gap, matrix = 124 × 120, field of view (FOV) = 220 × 220 × 125 (RL/AP/FH) mm$^3$, reconstruction= 256 × 256, diffusion direction = 3, NA = 1, parallel imaging with an acceleration factor of 2, and fat suppression (SPIR). Total acquisition time was 5 min 36 s.

Postprocessing

Brain extraction, motion correction and eddy current correction for DWI data were performed in FSL (FMRIB, Oxford, UK). All parametric maps were generated using an in-house MATLAB script (The MathWorks Inc., Natick, MA, USA).

Given that the relatively high diffusivity of CSF will cause fast signal attention with b-values, only b-values in the range of 0 and 3000 s/mm$^2$ were used for calculation. DKI analyses non-Gaussian water diffusivity using a polynomial model according to the following equation:(Jensen et al., 2005)

$$S_b = S_0 e^{\left\{-b*D_{app}+\frac{1}{6}*b^2*D_{app}^2*K_{app}\right\}}$$

where $S_b$ is the signal intensity at diffusion weighting factor b; $S_0$ is the signal intensity at b = 0 s/mm$^2$; $D_{app}$ is the apparent diffusion coefficient; $K_{app}$ is the apparent diffusion kurtosis, which is the indicator of the degree of non-Gaussian diffusion. The Levenberg-Marquardt (LM) algorithm was applied for optimization.

Due to the excellent contrast between brain parenchyma and CSF in voxelwise-computed diffusion-weighted imaging (vcDWI), it was calculated according to the following equation:(Gatidis et al., 2016)

$$vcDWI = S_0 e^{-D_{app}^2*10^6}$$

Representative mappings of the vcDWI and DKI metrics of CSF are shown in Figure 1.

Image analysis

In total, lateral cerebral ventricle (LV), the third ventricle (V3), and the fourth ventricle (V4) were analyzed. Regions of interest (ROIs) were manually drawn on continuous slices of vcDWI in ImageJ (NIH, USA). All ROIs were carefully placed to avoid parenchyma. The ROIs were delineated by a neuroradiologist (9 years of experience in neuroimaging diagnosis) and then, these ROIs were reviewed and confirmed by biomedical engineering with 14 years of experience. Then the ROIs of ventricles were projected to DKI parametric mappings. Mean, median, skewness and kurtosis of DKI metrics of each ventricle were extracted. The volume of ventricle was calculated as well.

Statistics

Statistical analysis was performed in MATLAB (The MathWorks Inc., Natick, MA, USA), MedCalc (MedCalc Software Ltd, Belgium) and SPSS (New York, USA). Data are presented as the mean ± standard deviation (SD). Normality was tested using the Kolmogorov–Smirnov test. Depending on the normality of the data, either the independent-samples t-test or the Wilcoxon rank-sum test were used for group comparison. First, demographic and clinical data were compared between the AD and HC groups. Given that significant between-group difference was found for age, P-values were corrected for age effect using logistic regression. Then, the associations between the DKI metrics and Mini-Mental State Examination (MMSE) scores of patients with AD and HC were investigated using partial correlation analysis with age, gender and education introduced as covariates.

Additionally, we also developed a classification model to differentiate patients with AD from HC based on DKI metrics, the volume of ventricle, age, gender and education. Only the normally distributed variables with P < 0.10 from the group comparison between AD and HC were considered for the classification modeling using Support Vector Machines (SVM), as described in previous study (Chen et al., 2020; Li et al., 2018). Then, feature selection was performed automatically using 10-fold SVM modeling with backward feather selection. The receiver operating characteristic curve (ROC) and the area under ROC (AUC) of individual selected feature was analyzed. For the predictive model using the combination of these four selected features, cross-validation was performed to avoid over-fitting. All datasets were split randomly and approximately equally into two groups (Groups A and B). Cross-validation was performed using these two groups. The optimal prediction model obtained in dataset A was validated on dataset B to evaluate its predictive performance, and vice versa. P<0.05 was considered to indicate a statistically significant difference.

Results

**Demography**

Table 1 tabulates the patient demographics and clinical information. The gender ratio showed no significant difference between the AD and HC groups (P>0.05). Among these included subjects, the AD group was significantly older than the HC group (P<0.001). HC groups showed a trend of higher education level, yet the difference was insignificant (P>0.05). As expected, MMSE score was significantly lower in the AD group than that in the HC group (P<0.001).

**Group comparison**

For all ventricles, the volume is significantly larger in AD than that in the HC group (LV=17.37$\pm$7.83 ml vs. 7.21$\pm$3.34 ml, corrected P=0.005; V3=1.29$\pm$0.46 ml vs.

$0.76 \pm 0.35$ ml, corrected P=0.004; V4=$1.82 \pm 0.66$ ml vs. $1.43 \pm 0.44$ ml, corrected P=0.039).

In previous literature using this method, non-physical kurtosis values were reported in brain parenchyma during the fitting. After the pixel-by-pixel check, no pixels in CSF showed non-physical kurtosis values in our work (Figure 2). This might be due to the relatively isotropic diffusion property of CSF.

After the correction using age as covariate, six DKI metrics showed significant differences (Figure 3). Median $K_{app}$ of LV and V3 were significantly higher in HC than that in the AD group (corrected P=0.025 and P=0.036, respectively), yet mean and median $K_{app}$ of V4 were significantly lower in HC than that in the AD group (corrected P=0.001 and P=0.004, respectively). Moreover, the skewness $D_{app}$ of LV and V3 differed significantly between these two groups (corrected P=0.011 and P=0.049, respectively).

**Correlation with the MMSE**

Age, gender and education level were considered as covariates in the analysis of the correlation between DKI metrics and MMSE. Among all analyzed parameters, no significant correlation was found in the AD group, while in the HC group, mean $D_{app}$ in LV (correlation coefficient=0.570, P=0.017), $D_{app}$ in V3 (correlation coefficient=0.497, P=0.042) and $D_{app}$ in V4 (correlation coefficient=0.510, P=0.037) showed significant correlation with MMSE.

**Cross-validated SVM predictive model**

The diagnostic power of DKI metrics of CSF and relative information for the differentiating between AD and HC was investigated. First, 10-fold SVM modeling using backward feature selection was performed. The volume of LV, mean $D_{app}$ of V3, and mean $K_{app}$ of V4/LV were automatically selected. The correlation matrix between the group category and these four selected features is presented in Figure 4. The ROC and AUC of individual selected feature are shown in Table 2 and Figure 5.

Second, we used these selected features to build a risk model using SVM. Since the sample sizes of these two groups are unbalanced, sample augmentation was applied(He et al., 2008). After sample augmentation, the sample size of the AD group remained the same (n=31), while 10 new samples were synthesized in the HC group (total n=30). To avoid over-fitting, cross-validation was performed. The results of cross-validation are shown in Figure 6 and Table 3.

In the predictive model, training with dataset A produced an accuracy of 96.77% (100.00% sensitivity and 93.75% specificity) with an AUC of 1.00 (95% CI: 0.98–1.00). The validation of this predictive model using dataset B enabled the correct prediction for 90.00% of the subjects (100.00% sensitivity and 100.00% specificity). The ROC demonstrated excellent predictive performance (AUC: 1.00; 95% CI: 1.00–1.00). The ROC curves are shown in Figure 6 (A). The other way round, training with dataset B produced an accuracy of 100.00% (100.00% sensitivity and 100.00% specificity) with an AUC of 1.00(95% CI: 1.00–1.00). The validation of this predictive model using dataset A enabled corrected prediction for 87.10% of the ROIs (100.00% sensitivity and 87.50% specificity) with an AUC of 0.96(95% CI: 0.89–1.00). The ROC curves are shown in Figure 6 (B). Compared with the AUC of a single variable listed in Table 2, the AUCs of SVM risk modeling were higher than that of any single variable.

**Discussion**

In this work, we used DKI to investigate the non-Gaussian diffusivity of CSF in AD group and HC group. Several DKI metrics showed significant between-group differences. The diagnostic ability of DKI metrics and related information was investigated and demonstrated excellent performance. It is worth noting that in HC group, after adjusting with age, sex and education level as covariates, there were significant correlations between DKI metrics and MMSE, but no significant correlation was found in AD group. This may indicate that the abnormal deviation of the correlation between the DKI metrics of CSF and MMSE may be used as an indicator of abnormal pathophysiology.

Most studies of AD based on magnetic resonance imaging have focused on the brain parenchyma(Belathur Suresh et al., 2018; Parker et al., 2018; Tardif et al., 2018; Worker et al., 2018), it is of interest to explore the MR characteristics of CSF since $A\beta$ and tau in CSF are well-accepted biomarkers of AD. Except for tau and $A\beta$, there are other chemical components in CSF, which can restrict the free-diffusion of water molecular as well(Falgàs et al., 2020). Moreover, it was reported that the tau level in CSF was significantly correlated with the atrophy rate of the hippocampal subfield in the converted mild cognitive impairment (c-MCI) group(Liu et al., 2021). Temperature can affect the diffusivity of water as well. Previously, Gianvincenzo Sparacia et al.(Sparacia et al., 2017) assessed the brain core temperature of AD patients in comparison with HC using diffusion-weighted imaging, and found slightly lower but insignificant difference in LV temperature (AD, 37.9 ± 1.1 °C; HC, 38.7 ± 1.4 °C, P=0.1937). Moreover, CSF is not a stationary solution, it flows slowly, the flow effect might affect the quantification of DKI metrics. IVIM-DKI model(Wurnig et al., 2016) can characterize the non-Gaussian diffusion property of CSF while taking the flow effect into account, but increased number of fitting parameters might decrease the accuracy of fitting or prolonged scan time is required to increase the SNR of the original DWI images.

In several previous studies(Tu et al., 2021; Xue et al., 2019), DKI has been explored for the diagnosis of AD, mainly focusing on the brain parenchyma. Min-Chien Tu et al. (Tu et al., 2021) investigated the discriminative ability of DKI metrics in segregated thalamic regions for the differentiating between subcortical ischemic vascular disease (SIVD), AD, and normal cognition (NC), and demonstrated a discriminant power of 97.4% correct classification. However, the brain of patient with AD often show atrophy and is accompanied by varying degrees of ventricular dilation, segmentation of brain regions and/or the subregions is problematic. In their work(Tu et al., 2021), double nonlinear registration was applied for segmentation, but the accuracy of segmentation is still a challenge, especially for patients with AD.

Moreover, the diagnostic power of other MR sequences was explored as well. Hao Guan et al. (Guan et al., 2022) proposed a structure MRI-based deep learning framework for AD diagnosis using the ADNI-1 and ADNI-2 dataset, and achieved mean AUC of 0.9265 and mean accurate prediction of 87.18% for cross validation. Buhari Ibrahim et al.(Ibrahim et al., 2021) conducted a systematic review aimed at evaluating the effect of functional connectivity (FC) of the default mode network (DMN) based on resting-state fMRI (rs-fMRI) in discriminating patients with AD and HC subjects, and achieved 81.25% sensitivity and 68.18% specificity. Dan Jin et al.(Jin et al., 2020) comprehensively explored one of the world's largest rs-fMRI biobank including data from six neuroimaging centers, with a total of 252 AD patients, 221 mild cognitive impairment (MCI) patients and 215 HC, and demonstrated mean AUC=0.85 for the cross-validated differential model.

Limitations

This study had several limitations. First, the sample size is relatively small. And the number of subjects in these two groups is unbalanced. Larger sample size is expected in the future. Second, this study was conducted at one single center. In order to develop universal prediction model, datasets from multiple centers are required. Thirdly, compared with brain regions, CSF is easy to be manually delineated, but personal bias cannot be completely avoided. In the future, automatic segmentation of lateral ventricles is expected.

In conclusion, the DKI metrics of CSF showed significant difference between AD and HC groups. The moderately significant correlation between $D_{app}$ in these three ventricles and MMSE in HC group might indicate the relationship between the balanced clearance of CSF and cognition. Moreover, cross-validated SVM predictive model based on DKI metrics and ventricle information demonstrated excellent performance.

## Acknowledgments

This work was supported by the Collaborative Innovation Program of Hefei Science Center, CAS (2020HSC-CIP010) and Collaborative Innovation Key Foundation of Hefei Science Center, CAS (2022HSC-CIP003).

## CRediT authorship contribution statement

Qiong Ye: study design, MR scan, post-processing, data analysis, interpretation, and writing the manuscript. Yingnan Xue: MR scan, data analysis and statistics. Min Wen: clinical information. All authors contributed to writing the manuscript.

## Conflict of interests

The authors have no financial or ethical conflicts of interest regarding the contents of this manuscript.

Table 1 Demographic and clinical data of the participants included in AD and HC groups.

|  | AD (n = 31) | HC (n = 20) | P value |
|---|---|---|---|
| Gender (male/female) | 11/20 | 8/12 | — |
| Age (year) | 64.9 ± 8.2 | 56.7 ± 6.3 | < 0.001 |
| Education (year) | 4.6 ± 4.3 | 6.5 ± 2.6 | 0.352 |
| MMSE | 18.5 ± 4.7 | 27.9 ± 1.6 | < 0.001 |

Abbreviations: AD, Alzheimer's disease; HC, healthy control; MMSE, Mini-Mental state Examination

Data are presented as mean ± standard deviation (SD).

Table 2. The AUCs of these four selected features.

| Variable | AUC | SE [a] | 95% CI [b] |
| --- | --- | --- | --- |
| **mean $D_{app}$ of V3** | 0.582 | 0.0742 | 0.448 to 0.707 |
| **volume of LV** | 0.953 | 0.0316 | 0.866 to 0.991 |
| **mean $K_{app}$ of V4** | 0.862 | 0.0488 | 0.750 to 0.937 |
| **mean $K_{app}$ of LV** | 0.692 | 0.0675 | 0.561 to 0.804 |

[a] DeLong et al., 1988

[b] Binomial exact

Table 3. The performance of cross-validated SVM classification based on four automatically selected features.

| | Training | | | | | Validation | | | |
|---|---|---|---|---|---|---|---|---|---|
| | **Accuracy** | **AUC(95% CI)** | **Specificity** | **Sensitivity** | | **Correct Prediction** | **AUC(95% CI)** | **Specificity** | **Sensitivity** |
| **A->B** | 96.77% | 1.00(0.98,1.00) | 100.00% | 93.75% | | 90.00% | 1.00(1.00,1.00) | 100.00% | 100.00% |
| **B->A** | 100.00% | 1.00(1.00,1.00) | 100.00% | 100.00% | | 87.10% | 0.96(0.89,1.00) | 100.00% | 87.50% |

Figure Legends

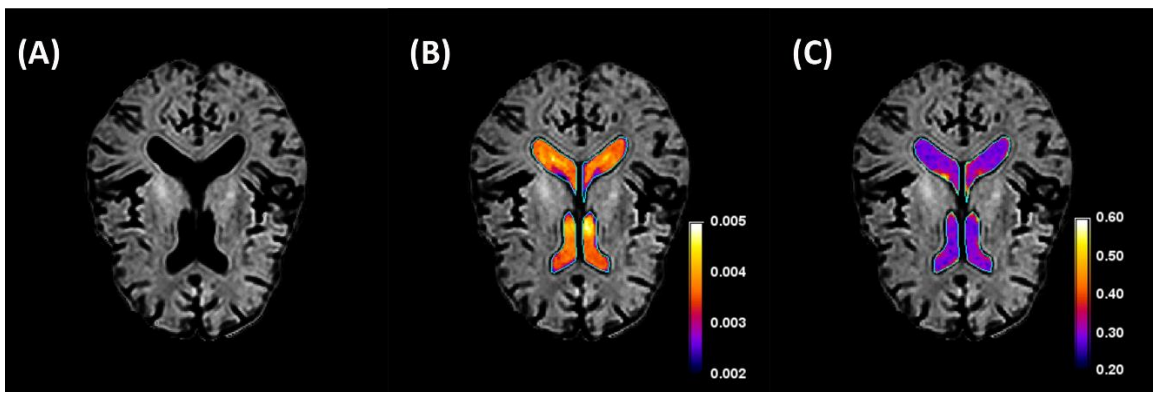

Fig 1. Representative vcDWI of patient with AD (A). The ROIs of LV are manually outlined. The overlap with $D_{app}$ (unit: mm$^2$/s) (B) and $K_{app}$ (unitless) (C) are displayed.

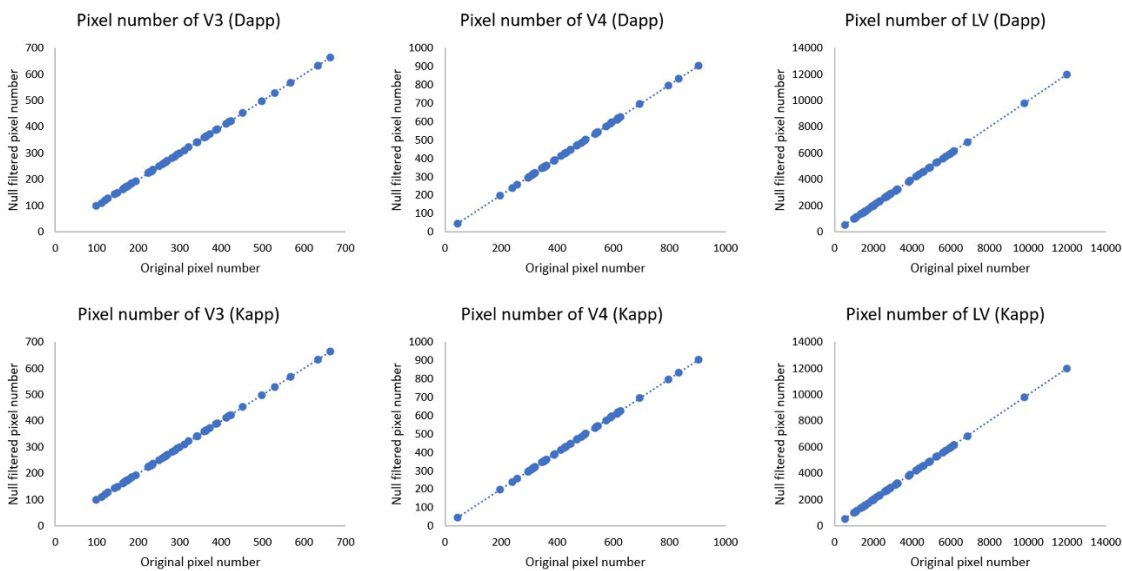

Fig 2. The plot of original pixel number of ventricles and that of null filtered.

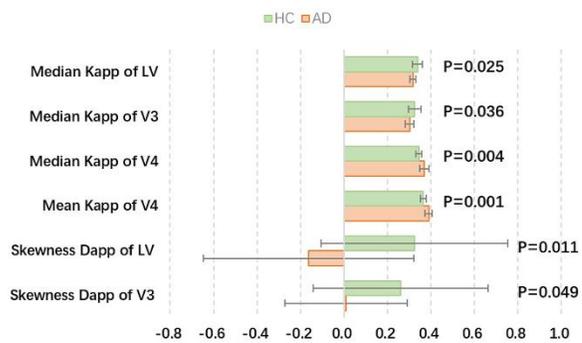

Fig 3. Group comparison of DKI metrics between AD and HC using age as covariate.

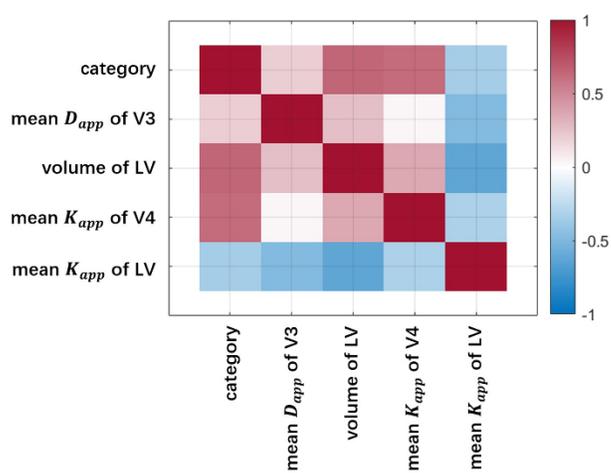

Fig 4. The correlation matrix between the group category and the selected features.

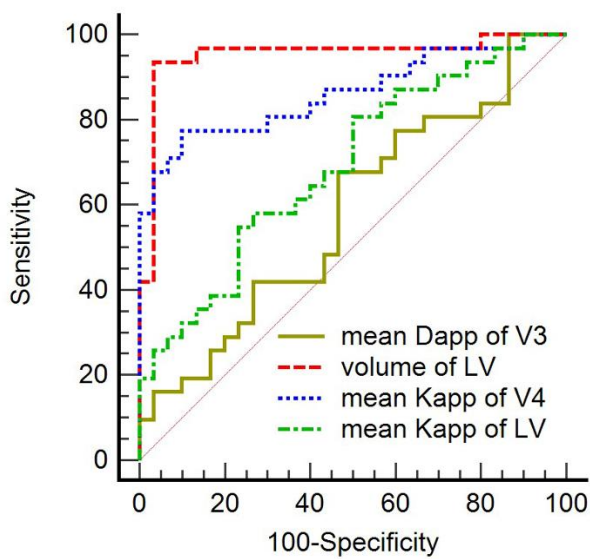

Fig 5. The ROC curves of these four selected features.

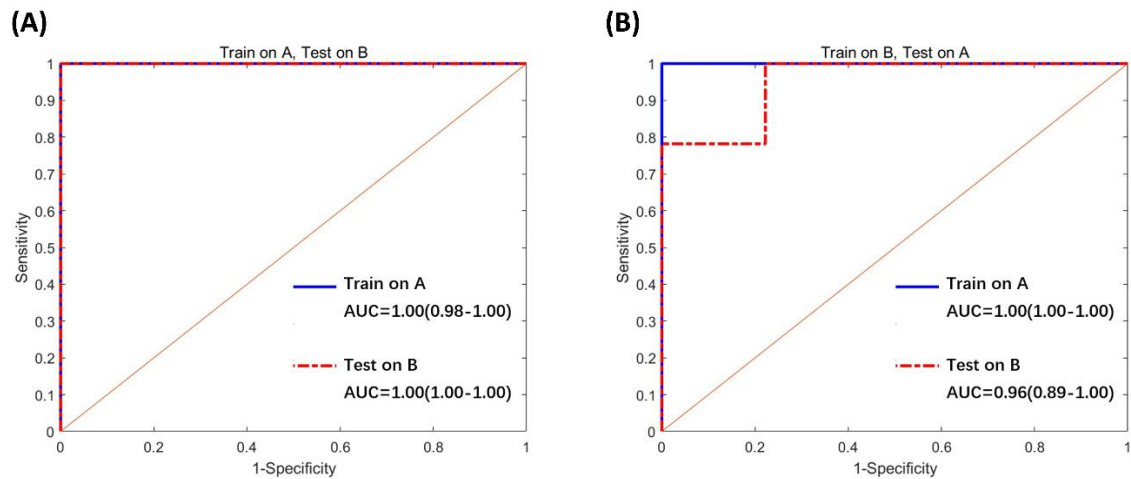

Fig 6. The performance of cross-validated SVM classification based on four automatically selected features is presented in ROC curves.